\documentclass[aps,prb,showpacs,twocolumn,superscriptaddress]{revtex4}

\usepackage[utf8]{inputenc}
\usepackage{amsmath}
\usepackage{amsfonts}
\usepackage{graphicx}
\usepackage{color}

\begin{document}

\title{Surface plasmons mediated energy transfer from a semiconductor quantum well to an organic overlayer}

\author{S. Kawka}

\affiliation{Scuola Normale Superiore, Piazza dei Cavalieri, 56126 Pisa, Italy}


\begin{abstract}
We consider the resonant energy transfer from a two-dimensional Wannier exciton (donor) to a Frenkel exciton of a molecular crystal overlayer (acceptor) when the active medias are separated by a metallic layer, possibly an electrode. We characterize the effect of the surface plasmon on this process. Using realistic values of material parameters, we show that it is possible to change the transfer rate within typically a factor of 5 (up to 44 according to geometrical configuration). We then take into account the quenching of the organic luminescence due to the proximity with the metal. This latter is significant and affect negatively the total internal efficiency that we discuss for different geometries.
\end{abstract}


\maketitle

\section{Introduction}

Semiconductor LEDs and their counterpart solar cells are expected to play a major role in general lightning and renewable energy production.
The development of organic light emitting diodes (OLED) has proven to be successful and they are on the market, notwithstanding the disadvantages of a poor carrier injection and transport properties compared to inorganic semiconductors. An idea to further improve such devices is the use of an hybrid system that would combine the best properties of both materials together and potentially open new possibilities for optoelectronic devices\cite{oulton}. In the strong coupling regime, novel hybrid quasiparticles are formed, with properties diverse from those of the individual excitons leading to new uses in particular those requiring large nonlinearities as for optical switching. In the weak coupling regime, in which Wannier and Frenkel excitons maintain their individuality, a suitable use of each component would also bring new features\cite{myrev} and could overcome the drawback presented earlier. For instance, carriers are electrically injected, transported and bound into excitons in the inorganic semiconductor subsystem, this latter being coupled to an organic light emitting subsystem via a F\"orster energy transfer process\cite{agranled}.
This internal non-radiative energy transfer had been proposed and has been recently demonstrated to be efficient enough (see  Ref.\onlinecite{agraCR} for a review of relevant work). Following early theoretical predictions\cite{agra97,agra99,denis,denisdot}, the energy transfer process has been observed  from  a quantum well  to a quantum dot overlayer\cite{klimov,zhang2007}, and from a quantum well to an organic overlayer\cite{heliotis2006,blumstengel2006,chanyawadee2008}.
A way to further improve the transfer between both components is the use of a plasmonic resonance to enhance the electric field in the system and take advantage of the metallic cathode which is often mandatory in electrical devices. Indeed the electrodes are often responsible of parasitic and strong waveguiding, a considerable energy is trapped in the surface plasmon mode then the recovery of some of this energy becomes then of interest\cite{an} 


We deal here with such hybrid systems structured in a planar geometry whereby a Wannier exciton in an inorganic semiconductor quantum well plays the role of the donor and a Frenkel exciton in a crystalline organic overlayer that of the acceptor. We study the role of an intermediate metallic layer that will carry the excitation through plasmons from the inorganic part to the organic. Moreover, we take into account retardation in the model and the possible quenching effects of the organic emission due to the metal layer.

In section \ref{model}, we describe the theory of  F\"orster energy transfer in a planar hybrid nanostructure taking into due account  plasmonic excitations and retardation. In section \ref{diel}, we use parameter values of typical hybrid structures to calculate the transfer rate for various configurations. We discuss the role of the plasmonic enhancement of the transfer and the additional loss channels introduced by the metal.

\section{Theoretical model}
\label{model}

\begin{figure}[!h]
\centering
\includegraphics[width=70mm]{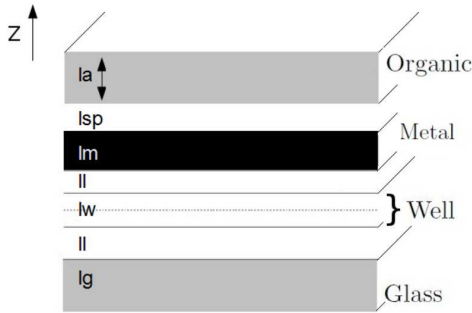}
\caption{Sketch of the planar hybrid heterostructure.}
\label{schema}
\end{figure}

We consider here the planar architecture shown in Fig.\ref{schema}. The donor subsystem consists of a semiconductor quantum well of thickness $l_w$ sandwiched between two semiconductor barriers of thickness $l_b$. For simplicity, we take the same background dielectric constant $\varepsilon_b$ for the well and the barriers and  we  assume the barriers to be infinitely high (i.e., the Wannier exciton is fully contained in the well region). At the bottom ($z<-l_b$) lies a transparent glass substrate with dielectric constant $\varepsilon_g$.  At the top lies a metallic layer described by his complex dielectric constant ${\varepsilon}_{m}$, a spacer layer with constant ${\varepsilon}_{sp}$ and then ($z>l_b+l_m+l_{sp}$) lies the acceptor subsystem consisting of a crystalline organic medium with complex dielectric constant ${\varepsilon}$ and thickness $l_a$ followed by air. The bottom glass and the air parts are supposed to be semi-infinite. The quantities $\varepsilon_g$, $\varepsilon_b$, $\varepsilon_{sp}$ and $\varepsilon_{air}$ include only the contribution of higher resonances (with respect to the exciton energies under consideration) and we consider them to be real. The quantity ${\varepsilon}_{a}$ is the total dielectric function of the organic material, including in particular the resonant absorption due to the Frenkel excitons, and thus it is a complex valued tensor \cite{kawka}, which however is assumed here isotropic for simplicity. The complex dielectric constant $\varepsilon_m$ also accounts for the metal losses.

Here we extend the theoretical framework for the F\"orster energy transfer rate in a hybrid nanostructure developed in references \onlinecite{agra97,myrev}, and discussed in detail in reference \onlinecite{denis}, to take into account retardation effects as well as plasmonic resonances. The former approach is equivalent to the usual F\"orster theory based on the dipole-dipole interaction \cite{agra} and leads to a macroscopic, semiclassical description of the energy transfer. Its main advantage is that the acceptor subsystem enters the calculation only via its complex (frequency dependent) dielectric tensor, a quantity directly accessible by experiment. In short, the transfer rate is obtained from the Joule losses suffered  in the organic medium by the electric field induced by the quantum well exciton, through the metal. This is appropriate to the weak coupling regime between donor and acceptor and occurs when the linewidth of the acceptor resonance is large compared to the interaction coupling donor and acceptor, leading to an irreversible energy transfer. Moreover, as the linewidth of 2D excitons in inorganic quantum wells are typically at least one order of magnitude smaller than the linewidth of the organic material excitons, we employ for simplicity a delta function approximation for the emission spectrum of the donor subsystem. As a matter of fact, the presence of the Wannier exciton of energy $E_{exc}=\hbar\omega$ gives rise to a source term in the quantum well corresponding to 
the exciton polarization oscillating at frequency $\omega$ which can be written as:
\begin{align} \label{pol}
\mathbf{P}(\mathbf{r})\,e^{-i\omega t}=\mathbf{d}^{vc} \psi(\mathbf{r},\mathbf{r})\,e^{-i\omega t},
\end{align}  
where $\mathbf{d}^{vc}$ is the matrix element of the electric dipole moment between the Bloch functions of the conduction and valence band extrema and $\psi(\mathbf{r}_e,\mathbf{r}_h)$ is the envelope function describing the bound electron-hole pair ($\mathbf{r}_e$, $\mathbf{r}_h$ being the electron and hole coordinates). 
In our simple model of a 2D Wannier-Mott exciton, the polarization is given by the product of the $1s$-wave function of the 2D relative motion of the electron-hole pair with the lowest subband envelope functions of electron and hole (which are equal in the approximation of infinitely deep well) and with the plane-wave function of the center-of-mass motion. This description is only appropriate to coherent large radius excitons with a well defined center of mass wavevector, as usually occur in high quality inorganic semiconductor quantum wells. Taking into account their corresponding normalization we can write:
\begin{align}\label{polarization}
\mathbf{P}(\mathbf{r})=\mathbf{d}^{vc} \sqrt{\frac{2}{\pi a_B^2}}\frac{1}{l_w}\cos^2\left(\frac{\pi z}{2l_w}\right)\frac{e^{i\mathbf{k}\mathbf{r}_\parallel}}{\sqrt{S}},
\end{align}  
where $S$ is the in-plane normalization area, $\mathbf{k}$ the in-plane wave vector of the center-of-mass motion, $\mathbf{r}_\parallel=(x,y)$ the in-plane component of $\mathbf{r}$ and $a_B$ is the 2D exciton Bohr radius.

Then, differently from previous work (Ref.\onlinecite{myrev},\onlinecite{agra97,agra99,denis,denisdot}), we look for the electric field $\mathbf{E}(\mathbf{r})\,e^{-i\omega t}$ resulting from this polarization calculated from Maxwell's equations with the use of transfer matrix theory \cite{savona}. 
The dielectric tensor $\varepsilon(z)$ is piecewise constant corresponding to each different layer, see Fig.\ref{schema}. We choose it to be complex in the organic layer and the metal to described the absorption and losses in these material, and real in the inorganic part, as the background absorption is supposed to be higher in energy, and for the rest of the system. 
The appropriate boundary conditions at the interfaces are the continuity of the tangential electric field and of the normal electric displacement between each layer.
Knowing the electric field, we can calculate the transfer rate (inverse transfer time) of the excitation into the organic medium:
 \begin{align}\label{taugene}
\Gamma_a=\frac{1}{\tau} = \frac{1}{2\pi\hbar} \int_{(\mathbf{r} \, \epsilon \, \mathrm{Org} )} \mathrm{Im}({\varepsilon}_{a}) \left(E_{a}(\mathbf{r})\right)^*E_a(\mathbf{r}) \, d^3r.
\end{align}  
This expression is equivalent to applying the Fermi Golden Rule to the decay of one excited state in the quantum well into the excited states of the organic molecules in the linear regime approximation (for a derivation of Eq.\ref{taugene} in terms of a microscopic inelastic scattering rate of Wannier excitons due to resonant two-level molecules in the organic medium see Ref.\onlinecite{denisigma}). In Eq.\ref{taugene}, and elsewhere below, the acceptor dielectric permittivity is evaluated at the frequency $\omega$ of the delta-like emission spectrum of the donor.
The power $W$ dissipated in the organic layer is given by $W=\hbar\omega/\tau$.
Such energy transfer mechanism, in the non-radiative approximation and without considering metallic layer, have been shown to be fast enough to efficiently quench the Wannier exciton luminescence and to turn on the organic molecule light emission \cite{denis}.

Due to the in-plane translational symmetry of the source (we are dealing with free excitons in the well), we consider the polarization for a given in-plane wave vector. In this case three modes of different symmetry can be identified: longitudinal (L) where $\mathbf{d}^{vc}$ is along the in-plane wave vector, perpendicular (Z) where the dipole moment is oriented along the z-axis, and transverse (T) for which the polarization is orthogonal to the two first cases. For each one, the exciton polarization (\ref{polarization}) gives rise to an electric field in the quantum well and then in all the structure. To calculate this latter, we determine for all three cases the source term to be included in the transfer matrix theory, as detailed below. 
With $q=\sqrt{\varepsilon(z) (\omega/c)^2-k^2}$ we can write for the TE mode :
\begin{align}
\nonumber E_T(z) &= A_+^s(z_0) e^{iq(z-z_0)} + A_-^s(z_0) e^{-iq(z-z_0)} \\
\nonumber B_L(z) &= \frac{cq}{\omega} \left[ -A_+^s(z_0) e^{iq(z-z_0)} + A_-^s(z_0) e^{-iq(z-z_0)} \right]  \\
B_Z(z) &= \frac{ck}{\omega} \left[ A_+^s(z_0) e^{iq(z-z_0)} + A_-^s(z_0) e^{-iq(z-z_0)} \right] 
\end{align}
 and the TM mode :
\begin{align}
\nonumber B_T(z) &= -A_+^p(z_0) e^{iq(z-z_0)} - A_-^p(z_0) e^{-iq(z-z_0)} \\
\nonumber E_L(z) &= \frac{cq}{\varepsilon \omega} \left[ -A_+^p(z_0) e^{iq(z-z_0)} + A_-^p(z_0) e^{-iq(z-z_0)} \right]  \\
E_Z(z) &= \frac{ck}{\varepsilon \omega} \left[ A_+^p(z_0) e^{iq(z-z_0)} + A_-^p(z_0) e^{-iq(z-z_0)} \right] 
\end{align}
with $s$ and $p$ the polarization.
The amplitude $A=\left( 
 A_+ , A_-  \right)$ from each side of a dielectric slab of thickness d are linked by $A(z_0 - d)=M_d A(z_0) $ with 
\begin{align}
M_d &= \left(
 \begin{array}{cc}
 e^{-iqd} & 0 \\
0 & e^{iqd} \end{array} \right)
\end{align}
At an interface between $\varepsilon_1$ et $\varepsilon_2$ we have  $A(z_0 - 0^+)=M_{\varepsilon_1 | \varepsilon_2}= A(z_0+0^+) $ with
\begin{align}
M_{\varepsilon_1 | \varepsilon_2}^s &= \frac{1}{2} \left(
 \begin{array}{cc}
 1+q_2/q_1 &  1-q_2/q_1 \\
 1-q_2/q_1 &  1+q_2/q_1 \end{array} \right) \\
M_{\varepsilon_1 | \varepsilon_2}^p &= \frac{1}{2} \left(
 \begin{array}{cc}
 1+\varepsilon_1q_2/\varepsilon_2q_1 &  1-\varepsilon_1q_2/\varepsilon_2q_1 \\
 1-\varepsilon_1q_2/\varepsilon_2q_1 &  1+\varepsilon_1q_2/\varepsilon_2q_1 \end{array} \right)
\end{align}
Considering a slab that include a source of field, as the quantum well considered here, a particular solution of Maxwell equations in the active domain leads to a source term $S$ for the amplitude. The relation of the amplitude at both ends is 
\begin{align}
A(z_0 - d-0^+)=M_{slab} A(z_0+0^+) + S\end{align}
where $S$ is related to the electric field at the border due to the intrinsic polarization distribution (\ref{pol}) and $M_{slab}$ is the matrix element for the whole slab, constituted by the matrix that we have previously seen, for a slab of thickness $d$ and dielectric permittivity $\varepsilon_2$ sandwiched between two medias of permittivity $\varepsilon_1$ and $\varepsilon_3$: $M_{slab}=M_{\varepsilon_1 | \varepsilon_2}M_{d}M_{\varepsilon_2 | \varepsilon_3}$ .

$S$ is given by the continuity relations of the fields at each interface. For $s$ and $p$ polarizations, writing  $z_-=z_0 - d+0^+$ and $z_+=z_0 - 0^+$ we obtain:
\begin{align}
S^s &= \frac{1}{2} \bigg[ \left(
 \begin{array}{c}
 E_T(z_-) \\
 E_T(z_-) \end{array} \right)  -M_{\varepsilon_1 | \varepsilon_2}^s M_{d}  \left(
 \begin{array}{c}
 E_T(z_+) \\
 E_T(z_+) \end{array} \right)  \bigg] \label{sources} \\
\nonumber
S^p &= \frac{1}{2} \bigg[ \left(
 \begin{array}{c}
\frac{\omega \varepsilon_2}{ck} E_Z(z_-) -  \frac{\omega \varepsilon_1}{ck} E_L(z_-) \\
\frac{\omega \varepsilon_2}{ck} E_Z(z_-) +  \frac{\omega \varepsilon_1}{ck} E_L(z_-)  \end{array} \right) \\
& -M_{\varepsilon_1 | \varepsilon_2}^p M_{d}  \left(
 \begin{array}{c}
\frac{\omega \varepsilon_2}{ck} E_Z(z_+) -  \frac{\omega \varepsilon_1}{ck} E_L(z_+) \\
\frac{\omega \varepsilon_2}{ck} E_Z(z_+) +  \frac{\omega \varepsilon_1}{ck} E_L(z_+)  \end{array} \right)  \bigg]
\label{sourcep}
\end{align}
where the fields $E_L$, $E_Z$, $E_T$ are the electric fields at the border of the active domain.

So starting from the polarization term  (\ref{polarization}), we solve Maxwell equations to find a particular solution in the quantum well domain. The field at the border will be used to find the general solution in the whole structure, giving a source term for the transfer matrix theory ((\ref{sources}) and (\ref{sourcep})). Choosing a symmetric particular solution according to the well width, the electric field is:
\begin{align}
E_w^{(L)} &= q^2 \frac{q_0^3 \sqrt{2} }{q^2 (q^2-q_0^2) \varepsilon_b \pi\sqrt{S}} \frac{d_L^{vc}}{a_B} \\
E_w^{(Z)} &=k ^2 \frac{q_0^3 \sqrt{2} }{q^2 (q^2-q_0^2) \varepsilon_b \pi\sqrt{S}} \frac{d_Z^{vc}}{a_B} \label{EwZ} \\
E_w^{(T)} &=\left(\frac{\omega}{c}\right)^2 \frac{q_0^3 \sqrt{2} }{q^2 (q^2-q_0^2) \varepsilon_b \pi\sqrt{S}} \frac{d_T^{vc}}{a_B}
\end{align}
with $q=\sqrt{\varepsilon (\omega/c)^2-k^2}$ and $ q_0={2\pi}/{l_w} $. That lead to the matrix elements (\ref{sources}) and (\ref{sourcep}) to be use:
\begin{align}
\mathcal{S}^s &=\frac{1}{2} \left(
\begin{array}{l}
1 -\cos({l_w q}) + i  \sin({l_w q})  \\  
 1 - \cos({l_w q}) - i  \sin({l_w q})  \end{array}
\right) E_w^{(T)} \\
\mathcal{S}^p &= \frac{1}{2} \left(
\begin{array}{l}
1 -\cos({l_w q}) + i \sin({l_w q})  \\  
 1 - \cos({l_w q}) - i \sin({l_w q})  \end{array}
\right) \frac{\omega \varepsilon_w}{c k}  E_w^{(Z)} \nonumber \\
 &+ \frac{1}{2} \left(
\begin{array}{l}
-1 + \cos({l_w q}) - i  \sin({l_w q})   \\  
 1 -   \cos({l_w q}) - i \sin({l_w q})  \end{array}
\right) \frac{\omega \varepsilon_w}{c q}  E_w^{(L)}
\end{align}
where the first row correspond to the traveling waves from the bottom to the top  in Fig.\ref{schema}, and the second row to the waves traveling in the opposite direction.
From here it is straightforward to determine the electric field in the whole structure and then the transfer rate to the organic media (\ref{taugene}). We also determine the losses in the process which compete with the energy transfer described by $\Gamma_a$. The main channel will be the dissipation in the metal $\Gamma_m$:
 \begin{align}\label{gammametal}
\Gamma_m=\frac{1}{2\pi\hbar} \int_{(\mathbf{r} \, \epsilon \, \mathrm{Metal} )} \mathrm{Im}({\varepsilon}_{m}) \left(E_{m}(\mathbf{r})\right)^*E_m(\mathbf{r}) \, d^3r.
\end{align}  
 and the direct radiative emissions $\Gamma_{rad}$ (as the final interest is the light emitted from the organic part) given by the flux of the Poynting vector in the $z-$direction in any section of the final layers, glass or air, as we consider them semi-infinite here:
 \begin{align}\label{gammaradl}
\Gamma_{rad;b,t}= \int_{S} \left( E(\mathbf{r}) \times B(\mathbf{r}) \right). \mathbf{e}_z \, d^2r.
\end{align}  
We also consider a non-radiative channel of dissipation $\Gamma_{nr}$ intrinsic to the well, independent of the $k-$vector, that will be estimated \cite{gladush} and is of the order a few hundreds of picoseconds.

We will discuss below the dependency with the $k-$vector, however in most cases of experimental interest, the transfer rate for an exciton distribution (rather than that for one specific exciton) should be considered via averaging over the initial state of the donor subsystem. We assume, in particular, that all directions for the center of mass wave vector of the exciton in the well are equiprobable (i.e., the exciton distribution has cylindrical symmetry). For a thermalized population of quantum well excitons, each one having a well defined center of mass wavevector, we also average over the energy $\frac{\hbar^2 k^2}{2M}$, with $M$ the exciton mass, according to the Boltzmann distribution, assuming that the wavevector dependent energy transfer itself is not fast enough to modify such exciton distribution. In the case of 2D excitons, the wave-vectors' amplitude distribution is given by:
\begin{align}
\rho (k)=\frac{ k e^{-\beta \frac{\hbar^2 k^2}{2M} } }{ \int e^{ -\beta \frac{\hbar^2 k^2}{2 m^*}} k dk} 
\label{distribT}
\end{align}
where $\beta=1/k_B T$. 
We also neglect here for simplicity the effects on the energy transfer efficiency of the dynamic ionization-recombination equilibrium between excitons and electron-hole unbound pairs which has recently been considered for GaAs based quantum wells \cite{gladush} and that should be less important in II-VI based quantum wells having a higher exciton binding energy.

After the excitation is transfered to the organic part, we have to consider the emission of this latter to evaluate the quenching due to the metal layer. We use the same approach as previously except a few details: as the Bohr radius of the Frenkel exciton is much smaller than the typical size of the layer, of the order of one molecule, we consider the emission from an assembly of point dipoles. The transfer by a point dipole at a given position $z_0$ {inside the organic layer} can be described using the preceding theoretical framework considering a sum over all $k$-vectors for an infinitely fine layer of emitters situated at {the same} position $z_0$. For each value of the $k$-vector the layer is defined by $\mathcal{S}_s^{org}$ and  $\mathcal{S}_p^{org}$:
\begin{align}
\mathcal{S}_s^{org} &= \left(
\begin{array}{c}
-1  \\  
 1  \end{array}
\right) \left( \frac{\omega}{c } \right)^2  {\frac{2 \pi i}{S}} \frac{d_T^{\cal{L}}}{q},\\
\mathcal{S}_p^{org} &= \left(
\begin{array}{c}
1 \\  
 1 \end{array}
\right) \left( \frac{\omega}{c } \right) \frac{2 \pi i}{S} d_L^{\cal{L}} \nonumber \\
 &+ \frac{1}{2} \left(
\begin{array}{c}
-1  \\  
 1 \end{array}
\right) \left( \frac{\omega}{c } \right) \frac{2 \pi i}{S} \frac{k d_Z^{\cal{L}}}{q},
\end{align}
{which are the limit of (\ref{sources}) and (\ref{sourcep}) when the width of the well is going to zero, with $1/a_B$ replaced by $1/\sqrt{S}$ for the point molecule, and $d^{vc}$ replaced by $d^{\cal{L}}=d^{vc}(\varepsilon+2)/3$ to take into account the local field effect.} We neglect the absorption due to the quantum well as the emission time for the organic is shorter than the transfer time to the quantum well, plus Stockes shift (weak coupling regime \cite{denis}).
From there, after integration, we obtain the electric field in the whole structure for a given molecule at position $z_0$ and then the radiative emission from the top $\Gamma_{rad;t}^{org}(z_0)$, and the losses: the dissipation in the metal $\Gamma_m^{org}(z_0)$ and the emission at the bottom in the glass side $\Gamma_{rad;b}^{org}(z_0)$. We integrate on all positions $z_0$ inside the organic to get the transfer rate from the assembly of molecules.

The initial excitation of each molecule is proportional to the energy transfered from the inorganic quantum well that reach its position. The detail of the spatial repartition of the energy in the organic before the reemission of that latter is strongly dependent of the material and is out of the scope of this article \cite{agra2012}. We consider here the two limiting cases of a weak and strong diffusion of the Frenkel exciton. If we can neglect the diffusion then the excitation is proportional to the square of the electric field at the point of the molecule (the integrand of eq.(\ref{taugene})), on the contrary for strong diffusion the energy spread in the whole layer and the excitation is independent of the position and proportional to $\Gamma_a$ (\ref{taugene}).

The plasmonic resonance will increase the transfer rate to the organic but also induce losses, that have to be evaluated so we detail the efficiency of the transfer $\eta_a$ and the efficiency of the organic emission $\eta_{org}$. This latter is different for weak and strong diffusion hypothesis. The two processes being independent, the internal efficiency $\eta_{int}$ will be the product of the two quantities. The total efficiency of the device, that take into account the out-coupling of the light and the injection process, will not be studied here. 
\begin{align}
\eta_a &= \frac{\Gamma_a}{\Gamma_a+\Gamma_m+\Gamma_{rad;b}+\Gamma_{rad;t} +\Gamma_{nr}}  \\
\eta_{org}^{no \, diff} &= \frac{ \int  \Gamma_{rad;t}^{org}(z_0) |E_a(z_0)|^2 dz_0 }{ \int \left( \Gamma_{m}^{org} +\Gamma_{rad;b}^{org} + \Gamma_{rad;t}^{org} \right) (z_0)   |E_a(z_0)|^2 dz_0 }  \\
\eta_{org}^{diff} &= \frac{ \int \Gamma_{rad;t}^{org}(z_0) dz_0 }{ \int \left( \Gamma_{m}^{org} +\Gamma_{rad;b}^{org} + \Gamma_{rad;t}^{org} \right) (z_0)  dz_0 }  \\
\eta_{int} &= \eta_{a} \eta_{org}  
\end{align}

The explicit analytic form of the obtained results is much more cumbersome than in previous theoretical work \cite{denis}, due to the more complex geometry including the metal layer and the retardation effects taken into account here. In the following we will then only discuss the results numerically.

\section{Numerical results and discussion}\label{diel}

\begin{figure}[!h]
\centering
\includegraphics[width=80mm]{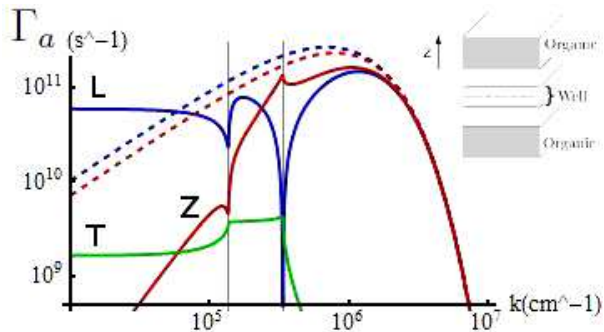}
\caption{Comparison of the model with retardation versus non-radiative, with the system and datas from Ref. \onlinecite{denis} and an emission at $2.7eV$ : plain lines are the exciton's transfer rate taking into account retardation effects for $L$-polarization (blue), $Z$-polarization (red), and $T$-polarization (green). Dashed lines are the exciton's transfer rate without retardation for $L$-polarization (blue) and $Z$-polarization (red). The geometry here consist of a ZnTe quantum well sandwiched between two layers of acceptor as in Ref. \onlinecite{denis}. The vertical lines are $\omega/c$ and $\sqrt{\varepsilon_g}\omega/c$ where $\varepsilon_g$ is the relative permittivity of the glass substrate.}
\label{compa}
\end{figure}

In Fig.\ref{compa} we show the results of Ref. \onlinecite{denis} (dashed lines) together with the transfer rate obtained with the full calculation including retardation effects described above (plain lines), for the different possible polarizations of the exciton at the energy emission of the ZnSe well $2.7eV$ and thickness of the organic layer of $40nm$. In the large $k$ values, where retardation effects can be neglected, we see that both model coincide. For small $k$ values corresponding to the radiative part the behavior is largely affected. The $L$-polarized exciton transfer rate show a plateau the value of which depends on the thickness of the organic layer, as the propagating modes penetrate all the layer. The $Z$-polarization transfer rate shows a similar behavior in the non-radiative regime, but goes to zero when going to normal incidence ($k=0$) this is a simple geometric effects as in this case there is no electric field at the interface (see eq.(\ref{EwZ})). We also notice a transfer from $T$-polarized exciton, mostly in the radiative range. This latter is null if we take only into account the non-radiative part, as in this case it does not give rise to any charge distribution nor electric field. In between these two regimes, the position and intensity of the peak is slightly modified and depends on the absorber thickness.

We turn now to the complete geometry of Fig.\ref{schema} including the metallic layer. For illustrative purposes, we will focus on the effects of the plasmons and thus only change the overlayer configuration keeping all the rest of the heterostructure fixed. We take into account a simplified description of the organic with only one exciton and we discuss only the  $L$-polarization state in the quantum well. The $Z$-polarization presents the same behavior in the non-radiative part of the spectrum, however the transfer is killed when we approach orthogonal incidence ($k$=0), and the transverse polarization has a low transfer, only noticeable in the intermediate region with no effect of the plasmon. {For the donor subsystem, we take a ZnTe semiconductor quantum well for which $\varepsilon_b \simeq 7.4$, $d^{vc}/a_B \simeq 0.1 e$ where $a_B$ is taken to be $28 \mathring{A}$ \cite{cingolani} (note that in a quantum well this value is half the one of the bulk) and $M=0.71m_0$  \cite{handbook} where $m_0$ is the electron mass.} For the structure, we take $l_w=60 \mathring{A}$, $l_b=40 \mathring{A}$, $l_m=400 \mathring{A}$, $l_{sp}=20 \mathring{A}$, $l_a=400 \mathring{A}$, while the glass substrate and the air top layer are semi-infinite in the z-direction\cite{kawka}. The organic medium is tetracene which absorption match the ZnTe emission\cite{kawka}. We assume that even close to the interface the power transferred to the organic medium does not lead to any saturation effect. Thus, the organic medium can be considered to be in the ground state and can be described by its linear dielectric response.
In simple cases, it is possible to use an effective model \cite{tavazzi08}, typically near the exciton resonances, where the optical response is modeled by a real background constant and several Lorentz transitions $j$, described by their energy $E_{j,0}$, coupling amplitude $A_j$ and damping $\gamma_j$:
\begin{equation}\label{ej}
\varepsilon_a(\omega)=\varepsilon_{\infty} + \frac{A_j\Gamma_j E_{j,0}}{E_{j,0}^2-(\hbar\omega)^2 - i\gamma_j \hbar\omega}
\end{equation}
$\varepsilon_{\infty} = 1.39$ is the background constant. The values of the parameters are obtained by fitting experimental data. For clarity in the discussion about the plasmonic effects, we restrict ourselves to the lowest excitation for which $E_0=2.38 eV$, $A=2.971$, $\gamma = 0.088 eV$ \cite{tavazzi08}.
For the metallic layer we use silver with a permittivity described by a combination of Drude and critical points models, as described in Ref. \onlinecite{laroche}, that cover the energy range $2-3eV$.

\begin{figure}[!h]
\centering
\begin{tabular}{lr}
 a) &\includegraphics[width=70mm]{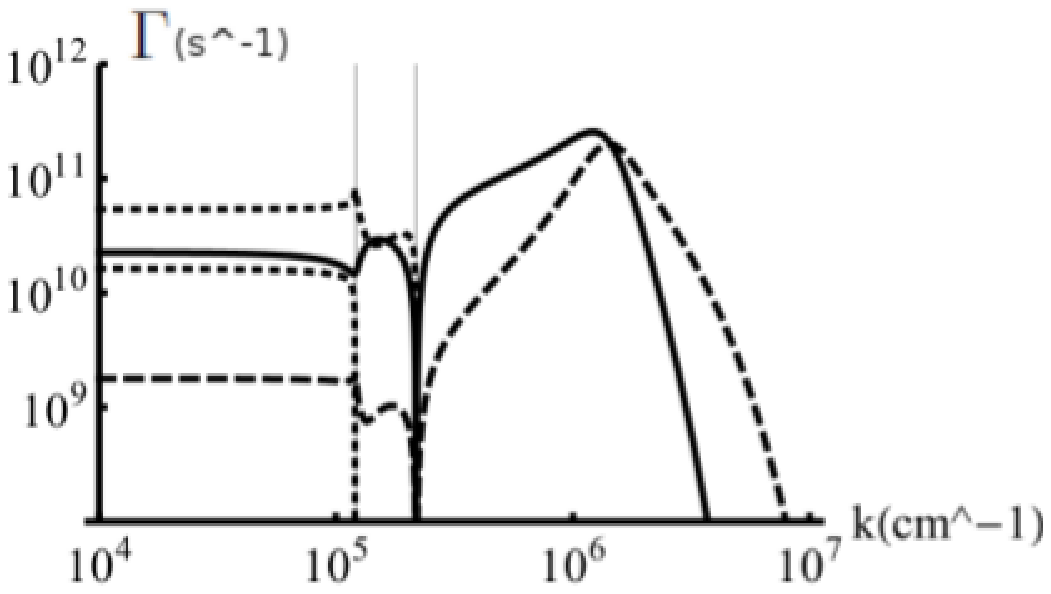} \\
 b) &\includegraphics[width=72mm,height=35mm]{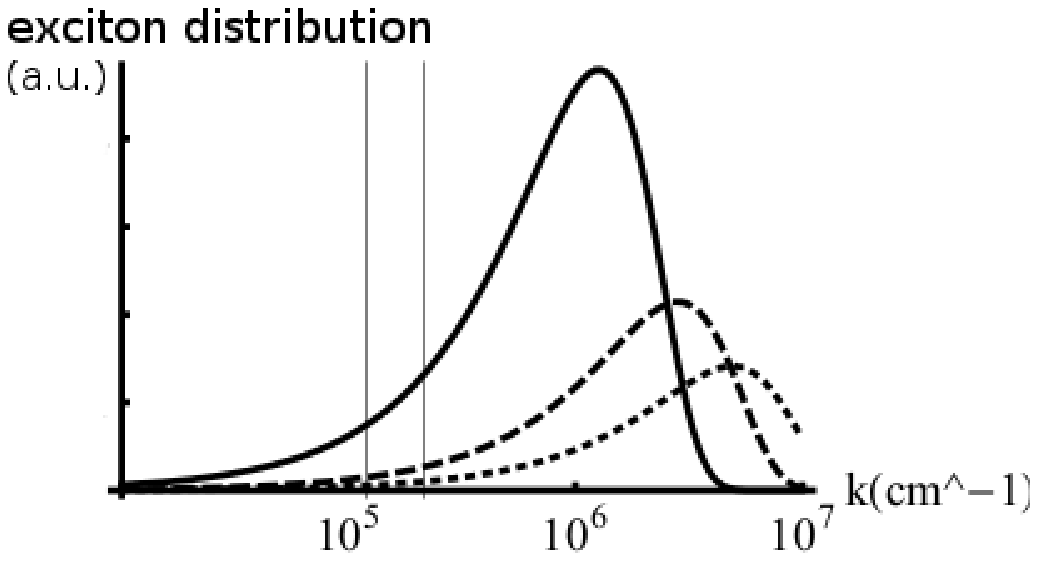}
\end{tabular}
\caption{a)Exciton's transfer rate for $L$-polarization to the organic part in presence of the metallic layer (plain line), losses induced in the metal (dashed) and trough radiative escape (dotted). The vertical lines are $\omega/c$ and $\sqrt{\varepsilon_g}\omega/c$ where $\varepsilon_g$ is the relative permittivity of the glass substrate.  b) Distribution of the exciton's $k-$vectors at $T=20K$ (plain line), $T=100K$ (dashed line), $T=300K$ (dotted line).}
\label{compametal}
\end{figure}
\begin{figure}[!h]
\centering
\includegraphics[width=80mm]{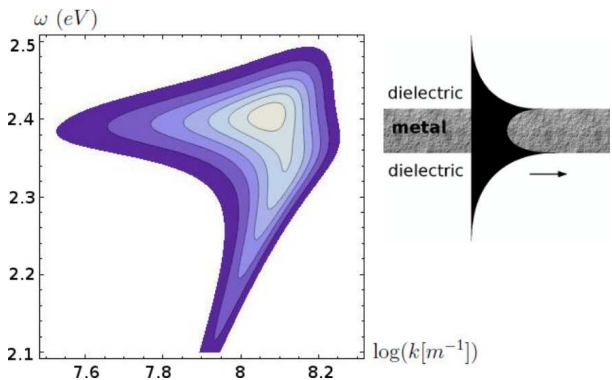}
\caption{Contour plot of the exciton's transfer rate for $L$-polarization to the organic part as a function of the $k-$vector and the frequency $\omega$, assuming that one can adjust the frequency of the quantum-well. The exciton's absorption lines, corresponding at $\omega=2.38$ cross the plasmons' lower branch, associated with the symmetric mode of the plasmon, sketched on the right.}
\label{3D}
\end{figure}

\begin{figure}[!h]
\centering
\includegraphics[width=80mm]{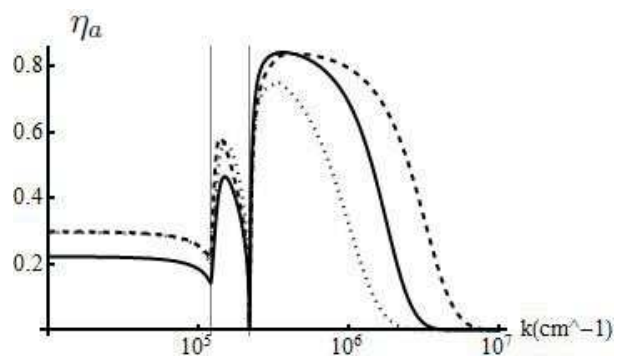}
\caption{Figure of merit $\eta_a$ of the exciton's transfer to the organic for $L$-polarization with the metal (plain line), without the metal (dashed) and without the metal keeping constant the gap between donor and acceptor (dotted).}
\label{Q}
\end{figure}

\begin{figure}
\centering
\includegraphics[width=65mm]{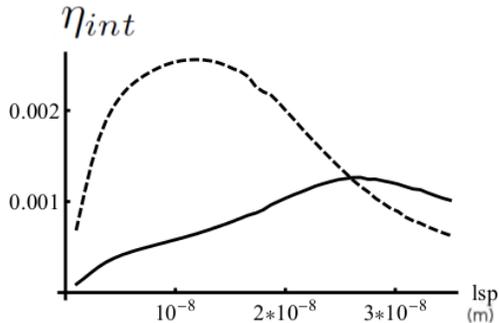}
\caption{Internal efficiency $\eta$ as a function of the spacer thickness, for an exciton at $T=300K$ with strong diffusion (dashed line) and weak diffusion (plain line).}
\label{QQ}
\end{figure}

In the following, we discuss the dependence of the transfer time on the $k-$vector in different geometries: system with the metal, without the metal but with the same distance between donor and acceptor, and without metal and spacer layer, and we evaluate the different loss channels. In Fig. \ref{compametal}a) we compare the transfer rates of the Wannier exciton to the organic part, to the metal and the radiative emission. The two vertical lines indicates the grazing incidence in the air and in the glass, thus the limit of the radiative (small $k$) and non-radiative (large $k$) domains at both extremities of the device. In the non-radiative part we distinguish the peak due to the plasmon resonance, that was not present in  Fig.\ref{compa}. The plasmon increases the local field (see sketch on Fig.\ref{3D}) and so the transfer to the organic, that lead also to a maximum of absorption by the metal as shown by the dashed curve. The dotted curves correspond to the light going out at both ends, in the air and in the glass. We remind that the abscissa is in a logarithmic scale, thus this latter case represent a small part of the spectra, moreover for a thermal distribution most of the excitons have a larger $k-$vector  Fig.\ref{compametal}b). At $T=20K$, $53\%$ of the energy is transfered to the organic, $43\%$ to the metal, $3\%$ is radiated and $1\%$ is lost in others internal non-radiative losses. Varying the (delta-like) donor frequency, we can see  in  Fig.\ref{3D} the absorption line of tetracene, the horizontal lines at energy $2.38eV$, the lower plasmonic branch and a small anti-crossing due to the coupling between the exciton and the plasmon. In this geometry we are in an intermediate regime between strong and weak coupling. The plasma frequency of silver being higher than the exciton line, only the symmetric mode of the plasmon participate to the transfer and is represented on the figure. 
\begin{figure}[!h]
\centering
\includegraphics[width=75mm]{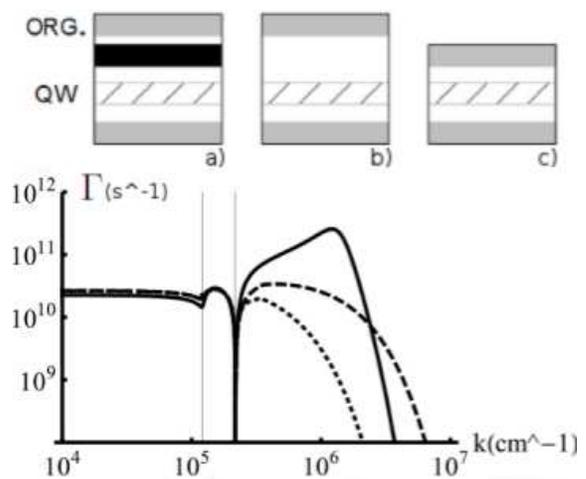}
\caption{Exciton's transfer rate for $L$-polarization to the organic part in the presence of the metallic layer (plain line, geometry a).), without the metal keeping constant the overall distance between donor and acceptor (dotted, geometry b).) and without the metal removing the layer  (dashed, geometry c).).}
\label{gammaconfig}
\end{figure}
\begin{figure}[!h]
\centering
\includegraphics[width=80mm]{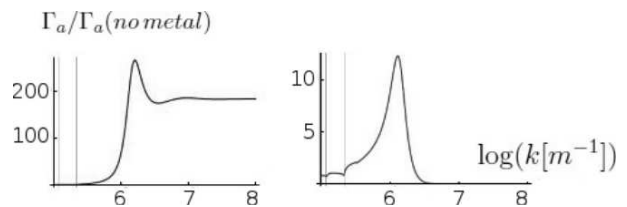}
\caption{$\Gamma_a / \Gamma_a (no \, metal)$ : comparison of the exciton's transfer to the organic for $L$-polarization with and without the metal keeping equal distance between donor and acceptor, as sketch in Fig.\ref{gammaconfig}b) (left) and without the metal removing the layer, as sketch in Fig.\ref{gammaconfig}c) (right).}
\label{Gammarapportssmet}
\end{figure}
We can see  Fig.\ref{gammaconfig} and \ref{Gammarapportssmet} that the coupling with plasmons increase significantly the transfer to the organic, up to 10 times for particular k-vectors when we add the metallic layer and to 200 times when we include the layer in an existing large barrier (respectively 5 and 44 times for a thermal distribution at 300K). The efficiency of this transfer is lower in the case where we add the metallic layer, due to the losses in the metal and to the larger distance between the quantum well and the organic, but is larger than the device with equivalent barrier, see  Fig.\ref{Q}. At 300K the efficiency of the transfer is respectively 35\%, 3.5\% and 28\% for the geometries a), b) and c) in Fig\ref{gammaconfig}. 
\begin{figure}[!h]
\centering
\begin{tabular}{ll}
 a) &  \includegraphics[width=70mm]{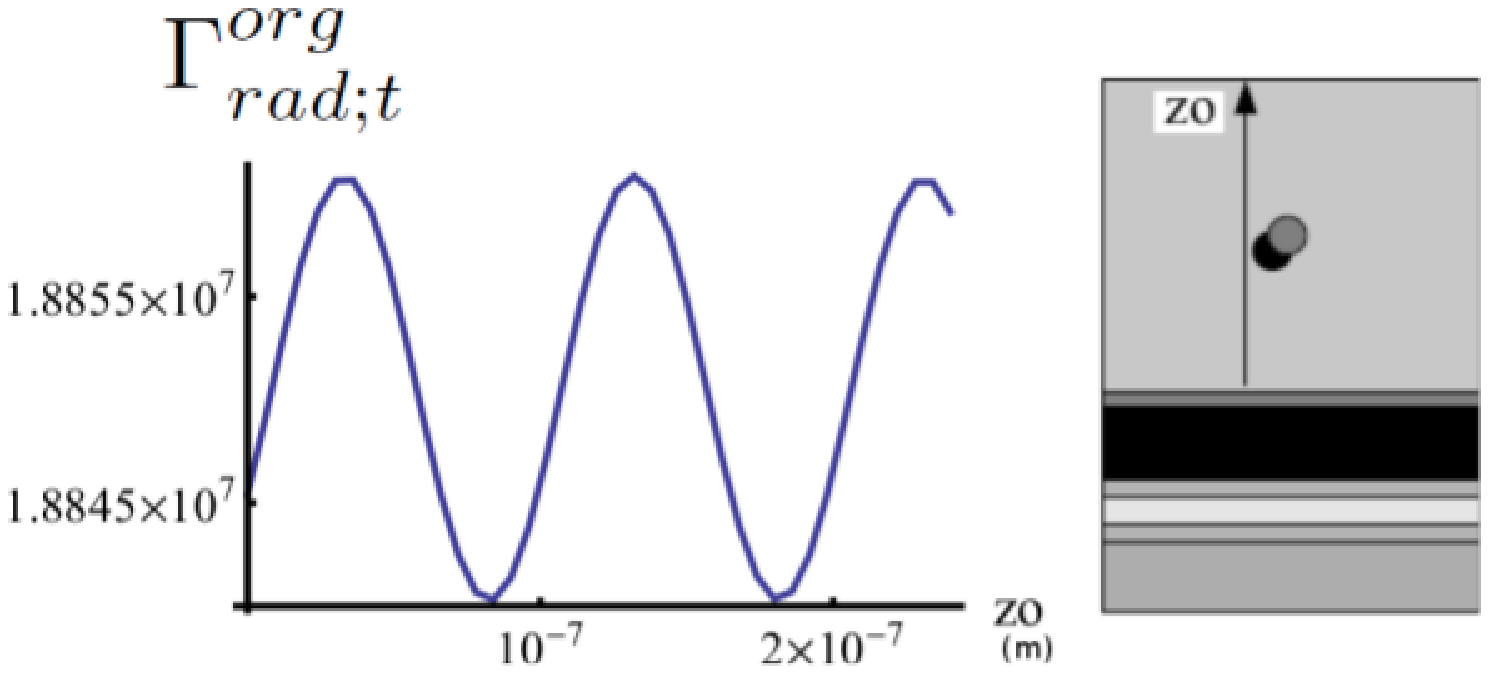} \\
 b) & \includegraphics[width=65mm]{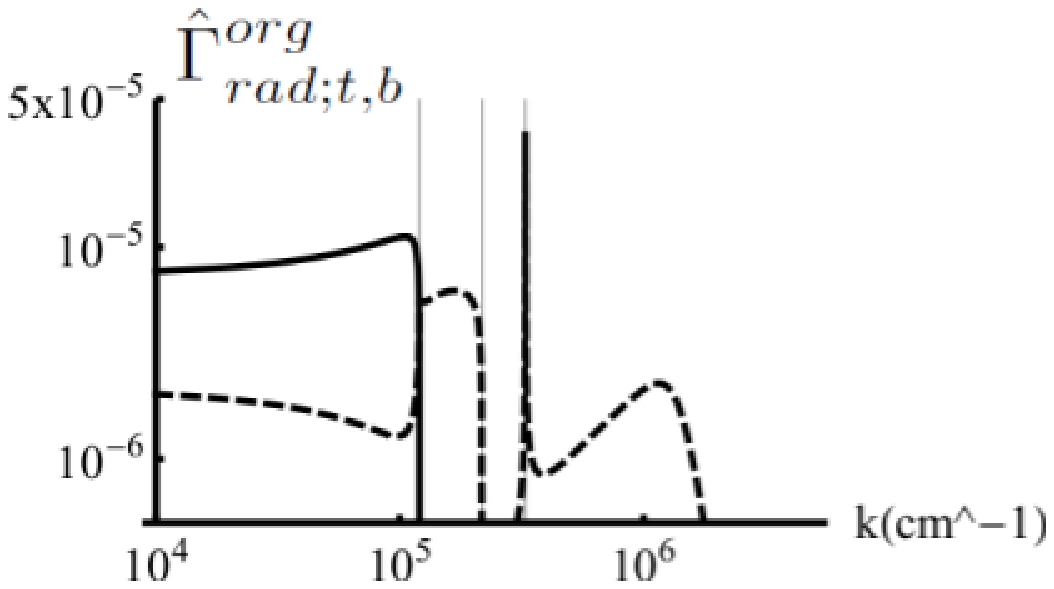}
\end{tabular}
\caption{a) Radiative emission to the air side of one molecule as a function of its position $z_0$ in the organic part, for a layer thickness of $240nm$. The oscillations are due the interferences with the reflected waves on the metal interface\cite{chance}.  b) {Emission of one molecule positioned at $z_0=l_a/2$ as a function of its Fourier component $k$, to the air side (plain line) and to the device side (dashed line) where we can see the emission to the semiconductor side and the absorption due the plasmon. The vertical lines are $\omega/c$, $\sqrt{\varepsilon_g}\omega/c$ and $\sqrt{\varepsilon_b}\omega/c$.}}
\label{onemol}
\end{figure}
The second part of the process is the radiative emission from the organic component, which can be negatively affected by the proximity with the metal layer. We consider each molecule as a single emitter. The radiative transfer in the open part of the device is depending on the emitter position due to interference effects (see Fig.\ref{onemol}) with an initial excitation density that depends on the position and on the exciton diffusion in the sample.  {After integration we find a radiative decay for the Frenkel exciton of about 40ns, which is in the range of the expected values for tetracene without considering any superradiance effect\cite{burdett}.} The spacing between the organic and the metal is a key parameter as it has an opposite effect on the transfer from the well and the radiative emission. 
At $T=300K$, the total efficiency is given Fig.\ref{QQ} and is less than one percent, 0.25\% in the case of a strong diffusion. That means that the quenching due to the presence of the metal is significant. 

\section{Concluding remarks}
We present a model for energy transfer in an hybrid device, taking into account both radiative and non-radiative part. We study the role of a plasmonic resonance due to the presence of an intermediate metallic electrode. We demonstrate that this latter is responsible of an effective increase of the internal energy transfer, up to a factor 5 to 44 at 300K according to the considered geometry, and more than a factor 200 for particular $k$-vectors. 
The plasmonic resonance leads also to a large quenching of the emission by the organic, which reduce significantly the final internal efficiency. A way to improve it further is to consider a patterned metal layer where most of the field will be directed in the dielectric, non absorbing holes in the metal, still taking advantage of the plasmonic enhancement.
\section*{ACKNOWLEDGMENTS}
The author would like to thank \hbox{G.C. La Rocca} for valuable discussions and for hosting this research, and acknowledges fruitful discussions with \hbox{V.M. Agranovich} and \hbox{D. Basko}.

This work is supported by the European FP7 ICARUS program, grant agreement no 237900.


\end{document}